\documentclass[twocolumn,prb,showpacs,floatfix]{revtex4}

\usepackage{graphicx}
\usepackage{bm}
\usepackage{color}
\usepackage{amsmath}
\usepackage{amsfonts}
\usepackage{amssymb}

\begin{document}

\preprint{APS/123-QED}

\title{Low-temperature thermal conductivity of BaFe$_2$As$_2$: Parent compound of iron-arsenide superconductors}

\author{N. Kurita$^{1}$} \author{F. Ronning$^1$} \author{C. F. Miclea$^1$} \author{E. D. Bauer$^1$} \author{J. D. Thompson$^1$} \author{A. S. Sefat$^2$} \author{M. A. McGuire$^2$} \author{B. C. Sales$^2$} \author{D. Mandrus$^2$} \author{R. Movshovich$^1$}
\affiliation{$^1$Condensed Matter and Thermal Physics, Los Alamos National Laboratory, Los Alamos, New Mexico 87545, USA}
\affiliation{$^2$Materials Science $\&$ Technology Division, Oak Ridge National Laboratory, Oak Ridge,
Tennessee 37831, USA}

\date{\today}

\begin{abstract}
We report low-temperature thermal conductivity down to 40\,mK of the antiferromagnet BaFe$_2$As$_{2}$,
which is the parent compound of recently discovered iron-based superconductors. In the investigated
temperature range below 4\,K, the thermal conductivity $\kappa$ is well described by the expression
$\kappa$\,=\,$aT$\,+\,$bT^{2.22}$. We attribute the ``$aT$''-term to an electronic contribution which is
found to satisfy the Wiedemann-Franz law in the $T$\,$\rightarrow$\,0\,K limit, and the remaining
thermal conductivity, $\sim$\,$T^{2.22}$, is attributed to phonon conductivity. A small influence on
thermal conductivity by magnetic fields up to 8\,T is well accounted by the observed magnetoresistance.
The result is consistent with a fully gapped magnon spectrum, inferred previously from inelastic neutron
scattering measurements.

\end{abstract}


\maketitle
\section{Introduction}
As in the high-$T_c$ cuprates, with electron- or hole-doping, or with application of external pressure,
superconductivity in the iron pnictides emerges upon suppression of the antiferromagnetism found in the
parent compounds\,\cite{KamiharaJACS2008,Rotter2008b,PinducedSC1,PinducedSC2}. As to their
superconductivity, in spite of a large number of experiments trying to determine the
structure of the superconducting gap, the question whether superconductivity in iron-pnictides is
conventional $s$-wave, unconventional $s$-wave or $d$-wave with node(s) remains a controversial
issue\,\cite{NakaiJPSJ2008,Hashimoto2008c,TYChenNature,DingEPL2008}. The detailed knowledge of the
antiferromagnetic state in the non-superconducting parent compound is a necessary ingredient to elucidate how superconductivity,
with transition temperatures which can reach
55\,K\,\cite{ZARen2008a}, emerges in these compounds. Soon after the discovery of superconductivity in
LaFeAs(O,F)\,\cite{KamiharaJACS2008}, it was established that the ground state of the
parent compounds was a metallic collinear antiferromagnetic
state\,\cite{Neutron1,Neutron2,Neutron3,Neutron4}. More recently the spin wave spectrum of several related
``122'' compounds, crystallizing in the tetragonal ThCr$_2$Si$_2$ structure, have been explored by
inelastic neutron scattering, revealing a steep spin wave dispersion with a gap of 6-10
meV\,\cite{SpinWave1,SpinWave2,SpinWave3}.

Low-temperature thermal conductivity is a sensitive probe of spin, charge, and lattice degrees of
freedom. In particular, magnetic excitations, such as spin waves mentioned above, can both carry heat
and scatter other excitations which transport heat, such as electrons and
phonons\,\cite{DouglassPR,Sales,Sologubenko,Smith,YamashitaNaturePhysics}. Furthermore, low temperature
thermal conductivity can be a powerful probe of the superconducting order parameter as recently
demonstrated in nodal superconductors Tl-2201\,\cite{Tl2201} and CePt$_3$Si\,\cite{IzawaCePt3Si},
multi-band MgB$_2$\,\cite{multibandswave}, and fully gapped  Ni-based arsenic superconductor
BaNi$_2$As$_{2}$\,\cite{KuritaPRL2009}. However, thermal conductivity studies also allow deduction of
electronic, phononic and magnetic contributions to the parent state out of which superconductivity
emerges. In this work, we report on the low temperature thermal conductivity of BaFe$_2$As$_{2}$, which
is the non-superconducting parent compound of recently discovered
hole-doped(Ba,K)Fe$_2$As$_{2}$\,\cite{Rotter2008b} and electron-doped
Ba(Fe,Co)$_2$As$_{2}$\,\cite{SefatPRL2008} superconductors. From the lack of magnetic field dependence,
we conclude that our measurements up to 4\,K are consistent with a gapped magnon spectrum. We also
extract the electronic and phononic thermal conductivity that will be useful for interpreting the data
of the doped compounds which become superconducting.

\section{Experimental Details}
Single crystalline BaFe$_2$As$_{2}$ was grown by a self-flux method described in
Ref.\,\onlinecite{SefatPRB2008BaFeNiAs2}, although not from the same batch as the one used in that
investigation. Thermal conductivity was measured by a standard one-heater and two-thermometers technique
on a plate-like crystal with dimensions of $\sim$\,1.5\,$\times$\,0.6\,$\times$\,0.1\,mm$^3$, with a
heat current $q$\,$\parallel$\,[100]. Pt wires spot-welded to the sample provided a thermal link to
heater, thermometers, and the bath. The heater and RuO$_2$ thermometers were thermally isolated from the
support frame by superconducting NbTi filaments which have a small thermal conductance at low
temperature ($\sim$\,10$^{-10}$\,W/K at 0.1\,K for each thermometer or heater). Electrical resistivity
was measured for electrical current $J$\,$\parallel$\,[100], using the same crystal with the same
electrical contacts as in thermal conductivity measurement. Thermal conductivity measurements were
performed down to 40\,mK and in magnetic fields up to 8\,T using a dilution refrigerator with a
superconducting magnet. For resistivity measurements we used a Quantum Design Physical Property
Measurement System. The field orientation was $H$\,$\parallel$\,[001].

\begin{figure}
\begin{center}
\includegraphics[width=0.95\linewidth]{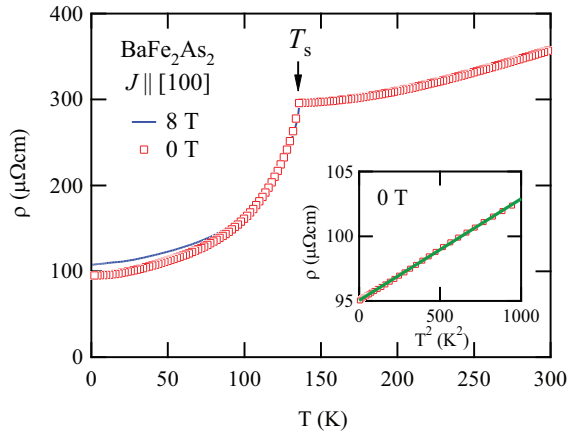}
\end{center}
\caption{(Color online) Temperature dependence of electrical resistivity $\rho$($T$) in zero field and
8\,T with current direction parallel to [100]. The arrow indicates the N\'eel temperature $T_\mathrm{N}$
coincident with the temperature $T_\mathrm{s}$ of the structural transition. The inset shows $\rho$ vs $T^2$. The
straight line, which was used to estimate an electronic thermal conductivity at low temperature, is a
least-square fit to $\rho$\,=\,$\rho_0$\,+\,$AT^2$} \label{fig1}
\end{figure}

\section{Results and Discussions}
Figure~\ref{fig1} shows the temperature dependence of electrical resistivity $\rho$($T$) of
BaFe$_2$As$_{2}$ in zero field and 8\,T. The data are consistent with previous
reports\,\cite{Rotter2008b,SefatPRB2008BaFeNiAs2}. An anomaly at $T$\,$\approx$\,140\,K appears due to
the simultaneous occurrence of a structural and magnetic phase transition~\cite{Rotter2008b}. As seen in
the inset, we fit the data at 0\,T below 30\,K to the Fermi liquid form $\rho$\,=\,$\rho_0$\,+\,$AT^2$,
where $\rho_0$ is the residual resistivity and $AT^2$ is ascribed to electron-electron scattering. The
fit, together with the thermal conductivity data, is employed below to test the Wiedemann-Franz (WF)
law. Using the value obtained $A$\,=\,0.0079\,$\mu$$\Omega$cm/K$^2$ and the reported electronic specific
heat coefficient $\gamma$\,=\,3\,mJ/Fe-mol\,K$^2$\,\cite{Rotter2008b,SefatPRB2008BaFeNiAs2}, we get a
ratio $A$/$\gamma^2$\,=\,8.8\,$\times$\,10$^{-4}$\,$\mu$$\Omega$cm/(mJ/mol\,K)$^2$ which is two order of
magnitude larger than the value of 1.0\,$\times$\,10$^{-5}$ observed in several heavy fermion compounds.
The origin of this enhanced ratio is unknown, but similar enhancement was also observed in other systems
which possess strong electronic correlations\,\cite{Kadowaki,MiyakeV2O3,NakatsujiCaSrRuO4}.

\begin{figure}
\begin{center}
\includegraphics[width=0.95\linewidth]{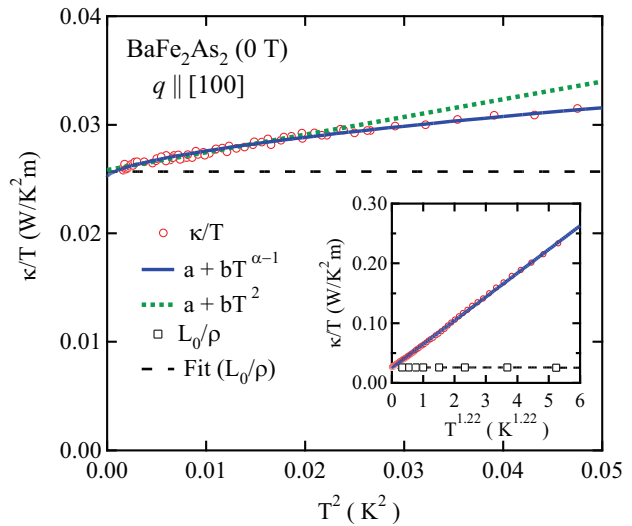}
\end{center}
\caption{(Color online)  Low temperature magnification of $\kappa$($T$)/$T$ as a function of $T^2$ in
BaFe$_2$As$_{2}$ for heat current $q$$\parallel$[100] in zero field. The inset shows $\kappa$/$T$ vs
$T^{1.22}$ in the temperature range up to 4\,K. Solid or dotted line represents a fit of the data to
$\kappa$\,=\,$aT$\,+\,$bT^\mathrm{\alpha}$ for $\mathrm{\alpha}$\,=\,2.22 or 3, respectively. Open
squares in the inset correspond to the electronic thermal conductivity $\kappa$$_e$\,=\,$L_0$$T$/$\rho$
with $L_0$\,=\,2.44\,$\times$\,10$^{-8}$\,W$\Omega$/K$^2$, derived from the resistivity data using the
Wiedemann-Franz law. Dashed line is the fit from the inset of Fig.~\ref{fig1} in thermal units.} \label{fig2}
\end{figure}

Next, we discuss the results of thermal conductivity measurements in BaFe$_2$As$_{2}$ down to
40\,mK. The thermal conductivity $\kappa$ can be expressed as the summation of multiple delocalized
excitations which can carry heat. In this system, we anticipate possible contributions from electrons,
phonons, and magnons, and thus
$\kappa\,=\,\kappa_\mathrm{el}\,+\,\kappa_\mathrm{ph}\,+\,\kappa_\mathrm{mag}$. From the magnetic field
dependence (shown below), we deduce that $\kappa_\mathrm{mag}\,=\,0$ in the investigated temperature
range below 4\,K. To separate the remaining electronic and phononic contributions, we rely on the fact
that the temperature dependence of the thermal conductivity for fermions and bosons is well known in
certain instances. In particular, when impurity scattering dominates the charge conduction, as
Fig.~\ref{fig1} clearly shows to be the case for our BaFe$_2$As$_{2}$ sample below 20\,K, it is
known\,\cite{Berman} that the thermal conductivity of electrons is proportional to $T$. The remaining
thermal conductivity can be assigned to phonons.

Figure~\ref{fig2} shows the $T^2$-dependence of $\kappa$/$T$ of BaFe$_2$As$_{2}$ in zero field.
 We use two different approaches to fitting the
low-temperature thermal conductivity data to the form $\kappa$/$T$\,=\,$a$\,+\,$bT^\mathrm{\alpha-1}$,
where $a T$ and $b T^\mathrm{\alpha}$ are electronic and phonon contributions, respectively. For
$\mathrm{\alpha}$\,=\,3, based on the conventional phonon scattering off the sample boundary (dotted
curve), we fit below 0.1\,K. In addition, we allow $\alpha$ to be a free parameter, (solid curve),
which results in a good fit over the entire temperature range measured (see Fig.~\ref{fig2}
inset). From these, we find $a$\,=\,0.0258\,W/m\,K$^2$, $b$\,=\,0.163\,W/m\,K$^4$ for the former and
$a$\,=\,0.0252\,W/m\,K$^2$, $b$\,=\,0.0396\,W/m\,K$^{3.22}$ and $\alpha$\,=\,2.22 for the latter. An electronic contribution
to thermal transport is govern by the Wiedemann-Franz (WF) law, which relates charge and thermal
conductivities by $\kappa$/$T$\,=\,$L_0$/$\rho$ with
$L_0$\,=\,2.44\,$\times$\,10$^{-8}$\,W$\Omega$/K$^2$. The values of $a$ for both fits to thermal
conductivity above are in excellent agreement with the expectation for the electronic contribution
$L_0$/$\rho$\,=\,0.0257\,W/m based on the value of $\rho_0$ obtained from the fit to the resistivity
data as indicated by a dashed line. Thus, we confirm that the WF law holds in BaFe$_2$As$_{2}$. The
remaining thermal conductivity is attributed to phonons and will be discussed in more detail below.

\begin{figure}
\begin{center}
\includegraphics[width=0.95\linewidth]{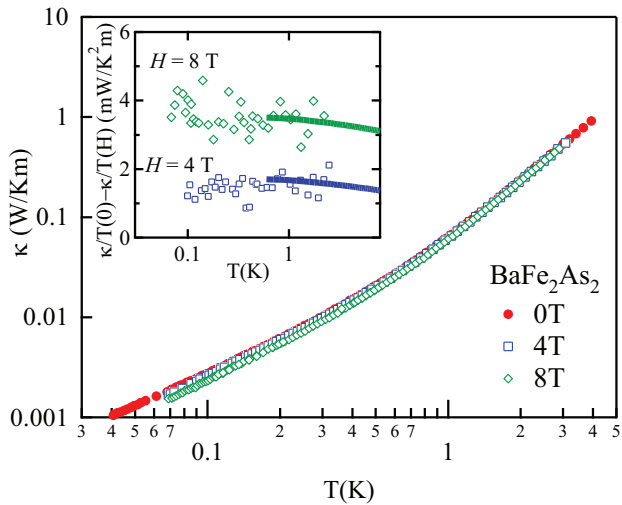}
\end{center}
\caption{(Color online) $\kappa$ vs $T$ at 0, 4 and 8\,T for the field direction $H$$\parallel$[001] and
$H$$\perp$$q$. The inset shows temperature dependence of $\Delta
\kappa/T$\,$=$\,[$\kappa/T(0)-\kappa/T(H)$] for $H$\,=\,4 and 8\,T. The bold solid curves represent
corresponding electronic term $\Delta L_0/\rho$\,$=$\,[$L_0/\rho(0)-L_0/\rho(H)$] derived from
measured resistivity data in fields.} \label{fig3}
\end{figure}

By applying a magnetic field up to 8\,T, we find a small suppression of the thermal conductivity at low
temperatures as can be seen in Fig.~\ref{fig3}. As shown in the inset, $\Delta
\kappa/T$\,$=$\,[$\kappa/T(0)-\kappa/T(H)$] exhibits almost constant negative values in the investigated
temperature and field range. In fact, the drop in thermal conductivity is fully accounted for by a drop
in electronic contribution, indicated by bold solid curves, due to magnetoresistance. The change in
electronic contribution is estimated via Wiedemann-Franz analysis from the resistivity data in
Fig.~\ref{fig1}. The remaining thermal conductivity, due to a combination of phonons and magnons, is
independent of magnetic field. Assuming that the phonon spectrum is field-independent  in
BaFe$_2$As$_{2}$ at low temperature, a non-superconducting material away from any structural or magnetic
instabilities, leads to a conclusion that the magnon contribution is also field independent, i.e.
magnetic excitations are gapped with $\Delta_\mathrm{mag}$\,$>$\,4\,K. The fully gapped magnon spectrum
we deduce is, hence, consistent with inelastic neutron scattering measurements that reveal that the spin
wave spectrum of BaFe$_2$As$_{2}$ has a 9.8\,meV gap\,\cite{SpinWave1}. Consequently, we would not
anticipate an effect from the magnons on the thermal conductivity until roughly 100\,K.

\begin{figure}
\begin{center}
\includegraphics[width=0.95\linewidth]{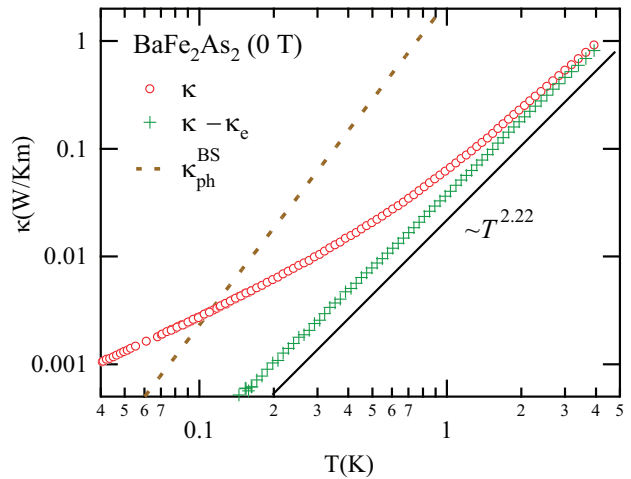}
\end{center}
\caption{(Color online) $\kappa$ vs $T$ in zero field and the extracted phonon term
$\kappa_\mathrm{ph}$\,=\,$\kappa$\,$-$\,$L_0$$T$/$\rho$ which follows $T^{2.22}$. The phonon thermal
conductivity based on boundary scattering $\kappa_\mathrm{ph}^\mathrm{BS}$ is shown with a dashed line.}
\label{fig4}
\end{figure}

We return finally to the thermal conductivity remaining after subtraction of the electronic
contribution, which we attribute to phonons $\kappa_\mathrm{ph}\,=\,\kappa\,-\,aT$. It satisfies a
single power law =$0.0396T^{2.22}$ over a large temperature range as can be seen in Fig.~\ref{fig4}. At
low temperatures, where these measurements are made, it is often found that a single scattering
mechanism is dominant and consequently the phonon thermal conductivity obeys a simple power law behavior
$\kappa_\mathrm{ph} = BT^\alpha$. For phonons scattering off the boundary of the crystal, one can
calculate the expected thermal conductivity using the formula\,\cite{Berman}:
\begin{eqnarray*}
\kappa_\mathrm{ph}^{\mathrm{BS}}\,=\,\frac{1}{3}C\langle v \rangle l_\mathrm{ph}
\end{eqnarray*}
where $C$($\propto$\,$T^{\alpha}$) is the heat capacity of the phonons per volume, $\langle$$v$$\rangle$
their velocity (both obtained from the experimentally measured heat capacity), and $\l_\mathrm{ph}$ is
an effective mean free path based on the crystal dimensions ($\l_\mathrm{ph}$\,=\,$\sqrt{4ab/\pi}$,
where $a$ and $b$ are sample cross section dimensions\,\cite{meanfreepath}). Since acoustic phonons have
a low temperature heat capacity proportional to $T^3$, one finds $\alpha$\,$=$\,3. If the faces of a
crystal are smooth, one can anticipate specular reflection of the phonons\,\cite{kph_Pohl} resulting in
a lower power law typically with $\alpha \approx 2.7$. We are not aware of a single scattering
mechanism, either theoretically or experimentally, over this temperature range which would give
$\alpha$\,$\sim$\,2.22. It should be noted that a similar magnitude of power law $\alpha$\,$\sim$\,2.4
has been reported in LiF after reduction of dislocation density by annealing\,\cite{kphAfterAnnealing}.

Phonons scattering off either grain boundaries or electrons are expected\,\cite{Berman} to give
$\alpha$\,=\,2, and so another possibility should be considered, that the observed power law temperature
dependence with $\alpha$\,=\,2.22 is not dominated by a single scattering mechanism, but is in fact a
combination of boundary scattering, phonon, and electron scattering.

Interestingly, a similar situation occurs in the cuprates\,\cite{kph_Taillefer}, where the low
temperature phonon thermal conductivity can be fit to the form
$\kappa_\mathrm{ph}$\,=\,$B$$T^\mathrm{\alpha}$ with $\alpha$ between 2 and 3. Taillefer and
coworkers\,\cite{kph_Taillefer} have argued that this empirical form provides for the most reliable
determination of the low temperature electron and magnon contributions to the thermal conductivity, and
suggest the origin of the $T^\alpha$ term is specular reflection off the smooth surfaces of the
crystals. On the other hand, Ando and coworkers maintain\,\cite{kph_Ando} that the obtained mean free
path is typically of the same order as the crystal dimensions which is not consistent with specular
reflection. Consequently, they attribute this strange power law simply to a crossover before the low
temperature boundary-limited scattering $T^3$ behavior is observed below $\approx$\,100\,mK.

In our measurements of BaFe$_2$As$_{2}$, the expected boundary-scattering limit
$\kappa_\mathrm{ph}^\mathrm{BS}$ is shown as a dashed line in Fig.~\ref{fig4}, using
$C$\,=\,8.98\,J/K$^{4}$\,m$^3$\,$\times$\,$T^3$ from Ref.\,\onlinecite{SefatPRB2008BaFeNiAs2},
$\langle$$v$$\rangle$\,=\,2400\,m/s and the
$\l_\mathrm{ph}$\,=\,$\sqrt{4ab/\pi}$\,=\,323\,$\mathrm{\mu}$m, where $a$\,=\,630\,$\mathrm{\mu}$m and
$b$\,=\,130\,$\mathrm{\mu}$m. The fact that it is larger than the measured $\kappa_\mathrm{ph}$
indicates our phonon mean free path is less than the sample dimensions, and continues to increase with
decreasing temperature, consistent with the power law exponent $\alpha < 3$. Thus, as in the cuprates,
we find a phonon thermal conductivity which is well described by a single power law with $\alpha < 3$
over two decades in temperature (40\,mK\,$\rightarrow$\,4\,K) and a mean free path slightly smaller than
the crystal dimension. The utility of this parameterization will be tested in future comparisons between
two families of high-temperature superconductors.

\section{Conclusion}

In conclusion, we have performed magneto-thermal conductivity experiments on the non-superconducting
antiferromagnet BaFe$_2$As$_{2}$, which has a high superconducting transition temperature by
electron- or hole-doping. The thermal conductivity $\kappa$ follows $\kappa$\,=\,$aT$\,+\,$bT^{2.22}$
in a wide temperature range below 4\,K. We attribute the ``$aT$''-term to an electronic contribution
which is consistent with the electronic conductivity expected on the basis of the Wiedemann-Franz law in
the $T$\,$\rightarrow$\,0\,K limit, and the remaining thermal conductivity, $\sim$\,$T^{2.22}$, is
attributed to phonon conductivity. A slight suppression of thermal conductivity by magnetic fields up to
8\,T corresponds to the observed positive magnetoresistance, and implies a fully gapped magnon spectrum
in BaFe$_2$As$_{2}$. This is an important step in understanding the low temperature thermal conductivity
of the doped compounds which become superconducting.

\section{Acknowledgement}

We would like to thank S.-H. Baek for useful discussions. Work at Los Alamos National Laboratory was
performed under the auspices of the US Department of Energy. Research sponsored by the Division of
Materials Sciences and Engineering, Office of Basic Energy Sciences.

\end{document}